\documentclass{article}
\usepackage[hyphens]{url}
\usepackage[utf8]{inputenc}
\usepackage{acl}
\usepackage{comment}
\usepackage{graphicx}

\usepackage{times}
\usepackage{latexsym}
\usepackage{comment}
\usepackage{multirow}
\usepackage{booktabs}
\usepackage{lscape}
\usepackage{rotating}

\usepackage{enumitem}
\setitemize{leftmargin=*,noitemsep}

\usepackage[T1]{fontenc}

\usepackage[utf8]{inputenc}

\newcommand\blfootnote[1]{%
  \begingroup
  \renewcommand\thefootnote{}\footnote{#1}%
  \addtocounter{footnote}{-1}%
  \endgroup
}

\title{A Comprehensive Framework to Operationalize Social Stereotypes \\for Responsible AI Evaluations}

\author{
  Aida Davani* \\
  Google Research \\
  \texttt{aidamd@google.com} \\\And
  Sunipa Dev* \\
  Google Research\\
  \texttt{sunipadev@google.com} \\\AND
  Héctor Pérez-Urbina \\
  Google Research \\ 
  \texttt{hekanibru@google.com}\\ \And
  Vinodkumar Prabhakaran \\
  Google Research \\ 
  \texttt{vinodkpg@google.com}
  }
\date{}

\begin{document}

\maketitle

\begin{abstract}
Societal stereotypes are at the center of a myriad of responsible AI interventions targeted at reducing the generation and propagation of potentially harmful outcomes. While these efforts are much needed, 
they tend to be fragmented and often address different parts of the issue without adopting a unified or holistic approach to social stereotypes and how they impact various parts of the machine learning pipeline. As a result, current interventions fail to capitalize on the underlying mechanisms that are common across different types of stereotypes, and to anchor on particular aspects that are relevant in certain cases. 
In this paper, we draw on social psychological research and build on NLP data and methods, to propose a unified framework to operationalize stereotypes in generative AI evaluations.
Our framework identifies key components of stereotypes that are crucial in AI evaluation, including the target group, associated attribute, relationship characteristics, perceiving group, and context. We also provide considerations and recommendations for its responsible use.
 
 \textcolor{red}{\textit{CONTENT WARNING: This paper contains examples of stereotypes that may be offensive.}}
 
\end{abstract}

\section{Introduction \& Motivation}
\blfootnote{* equal contribution}

Recent years have seen unprecedented gains in generative AI models' capabilities across modalities---language \cite{anil2023palm,achiam2023gpt}, image \cite{rombach2022high,saharia2022photorealistic}, audio \cite{kreuk2022audiogen,borsos2023audiolm}, and video \cite{ho2022imagen,bar2024lumiere}---while simultaneously gaining traction in diverse application domains and usage contexts across the globe \cite{sengar2024generative,raiaan2024review}. Along with these advancements, there are growing concerns that these models may reflect, propagate, and amplify societal stereotypes in their predictions and generations \cite{garg2018word,blodgett2020language,dev2022measures,hovy2021five}, potentially leading to downstream harms \cite{field2021survey,shelby2023harms}.

A growing body of empirical work shows how NLP models reflect societal stereotypes about various groups---including gender \cite{bolukbasi2016man}, race \cite{sap2019risk}, nationality \cite{jha-etal-2023-seegull}, and disability \cite{hutchinson2020social} to cite a few. 
Many efforts also build datasets to enable large-scale evaluation of stereotypes in model predictions \cite{nadeem2021stereoset,jha-etal-2023-seegull,bhutani-etal-2024-seegull}.
However, current research and resources
lack a unified approach toward stereotypes in AI, hindering a comprehensive understanding of the problem space and, thereby, limiting effective and scalable interventions. 
First, they fail to capitalize on the underlying common mechanisms that may be contributing to stereotypes in society, data, and models. Consequently, it makes it harder to envision a unified way to tackle and prioritize downstream sociotechnical harms; which could instead lead to unintended consequences, like new stereotypes emerging when others are mitigated.
Another gap stems from adopting simplistic representations of stereotypes for expediency in evaluations, e.g., \textit{(identity, attribute)} pairs overlook core aspects such as how stereotypes tie to specific time and place, which social groups hold certain stereotypes, and what connotations they imply. 

Finally, there are different methodologies to source stereotype data---e.g., annotator-driven collection \cite{nadeem2021stereoset}, LLM-enabled collection \cite{jha-etal-2023-seegull}, and community-centered collection \cite{dev-2023-building}---each having unique strengths in terms of scalability, coverage, and reliability. However, we currently do not have an effective approach to determine which of these methods are appropriate in which contexts, what their relative merits (and demerits) are, and how to use these approaches in ways that lean on their strengths and complementarities. 
Having a unified framework will enable effective intervention, prioritization in high-stake environments, shared knowledge and methods across various efforts to collect data and intervene on models, predictions, and evaluations. Such a framework will also reveal aspects of this problem space that we still have large gaps to fill.

In order to address these needs, we build off of social scientific theories on stereotypes as well as existing research on evaluating language technologies for stereotypes, and propose a unified, comprehensive framework to operationalize stereotype evaluations. Our framework identifies various high level components such as the \textbf{target group}, the \textbf{attribute} associated with the group, the characteristics of their \textbf{association}, the \textbf{perceiving group}, as well as the \textbf{context} within which these stereotypes are prevalent. We also outline a set of  recommendations for how to factor in responsibility considerations while using this framework.

\section{Background}

Social scientists have dedicated substantial research to the study of stereotypes, recognizing their intricate and multifaceted nature \cite{macrae1996stereotypes,schneider2005psychology}. This exploration has led to the development of various frameworks over time, aiming to unravel the complexities of how stereotypes originate, function, and influence both individuals and society as a whole \cite{hilton1996stereotypes}. Early work predominantly viewed stereotypes as inaccurate generalizations about groups, stemming from limited or biased information \cite{allport1954nature}. Stereotypes are also seen as cognitive shortcuts that help individuals simplify and categorize the social world, although this simplification could lead to errors and biases \cite{dovidio2010prejudice}. While these cognitive processes can be efficient, 
the connection between stereotypes (cognitive bias), prejudice (attitude bias), and discrimination (behavioral bias) was recognized early on, pointing to stereotypes as the motivation for negative attitudes and behaviors toward out-groups \cite{macrae2000social}.

Various theories have been developed that focus on diverse aspects of stereotypes. The \textit{Social identity theory} emphasizes the role of group membership in shaping self-concept and inter-group relations, suggesting that stereotypes can serve to enhance one's own group identity \cite{tajfel1979integrative}. The \textit{Social learning theory}, on the other hand, focuses on stereotypes being learned through observation and socialization, often from parents, peers, and media \cite{bandura1977social}. The \textit{System justification theory} examines how stereotypes can be used to justify existing social hierarchies, even by members of disadvantaged groups \cite{jost1994role}. Finally, \textit{Intersectionality theory} further emphasizes the interconnected nature of social identities and how multiple stereotypes can intersect to create unique experiences of discrimination \cite{crenshaw2013demarginalizing}. 

These theoretical perspectives have guided the development of various frameworks for analyzing stereotypes. Primarily shaped by social psychologists, these frameworks are widely used in other fields to model group dynamics and interactions. 
One of the prominent such frameworks is the \textit{Stereotype Content Model} (SCM), which posits that stereotypes vary along two dimensions: \textbf{Warmth} and \textbf{Competence}, resulting in different emotional and behavioral responses toward groups \cite{cuddy2007bias,fiske2018model}. By extending the SCM, the \textit{dual perspectives model} \citep{abele2016facets} added Morality and Sociability axes to the Warmth, and Ability and Assertiveness axes to the Competence dimension. 
\textit{Agency-Beliefs-Communion} \cite[ABC;][]{koch2016abc} model further added Status to the Competence dimension and  Belief as a dimension; specifically, ``one end of Beliefs represents all religious, conservative, and other traditional groups; at the other end are progressives, artists, scientists, and LGBTQ groups.'' \citet{nicolas2022spontaneous} relied on natural language processing approaches to both validate the SCM's dimensions as well as discovering dimensions not commonly covered by SCM, such as Health and Appearance. 

Some of these frameworks are increasingly being explored in NLP research. For instance, SCM has been applied to understand annotator biases \cite{davani2023hate} and debiasing word embeddings \cite{ungless2022robust,omrani-etal-2023-social}. \citet{fraser2022computational} present a computational method to apply SCM to textual data and demonstrated that stereotypes in textual resources compare favorably with survey-based studies in the psychological literature.
\citet{fraser2024does} used the ABC dimensions to evaluate and compare biases toward occupational groups across traditional survey-based data and various text sources. 
As NLP efforts increasingly grapple with the complexities of stereotypes in language, relying solely on social psychological frameworks of stereotypes can limit the scope of the analyses.  These frameworks often prioritize dimensions like warmth and competence, potentially overlooking crucial aspects such as social dynamics, socio-historical context, and linguistic valence, which are also essential for a comprehensive understanding of stereotypes in language technologies.

\section{Reflective exercise} 

In this section, we present a reflective exercise on NLP research on social stereotypes with the objective of demonstrating various focus areas surrounding this topic. For comprehensive surveys on this active research area, see \citet{blodgett2020language,blodgett-etal-2021-stereotyping}. 

\subsection{Stereotype Detection and Evaluation}

A significant number of responsible AI and NLP evaluations are concerned with various concepts that are inherently intertwined with stereotypes. For instance, bias measurement in co-reference resolution tasks often relies on gender-based occupation stereotypes \cite{zhao-etal-2018-gender,rudinger-etal-2018-gender}; hate speech detection can hinge on societal stereotypes \cite{chiril-etal-2021-nice-wife}; offensive text can be comprised of stereotypes~\cite{jeong-etal-2022-kold}; sentiments that are disparately associated with different target groups stem from stereotypical perceptions about them~\cite{kiritchenko-mohammad-2018-examining}; and more. 
However, the stereotype resources that these evaluations depend on, are limited in which groups they represent. While substantial work has focused on gender and racial stereotypes, they are also mostly constrained by binary gender constructs~\cite{dev-etal-2021-harms} and Western racial histories~\cite{sambasivan2021re}. Other identity axes such as disability status or socio-economic conditions are not as well represented. These resources are also rife with Western gaze wherein a majority of the resources are collected in the West (or even specifically North America), with data and annotators both representing Western viewpoints.

Based on keyword-based querying of the ACLanthology,\footnote{https://aclanthology.org/} we note that 4140 papers mention stereotypes, their detection, resources, and evaluation. Of these, 54.1\% mention gender-based stereotypes, 25.8\% mention racial stereotypes, and only 16.4\% mention region- and nationality-based stereotypes, and an even smaller fraction mention other identities such as age, disability, and profession. 
Some papers categorize stereotypes as positive or negative, often discussing the associated sentiment rather than the effect it can have downstream or the specific marginalization the target groups experience \cite{blodgett-etal-2021-stereotyping}. For example, ``women are polite'' can arguably be considered positive because of the sentiment associated with politeness, but the stereotype can have other implicit harms~\cite{cheng-etal-2023-marked} related to the history of expectations of politeness and servitude from women~\cite{garg-etal-2018-word}, something that can negatively influence applications such as job recommendations based on gender.

\subsection{Stereotype Resource Creation}
\label{sec:collecting} 
Evaluating how stereotypes impact NLP model outputs requires societal data that capture such stereotypes. In this section, we discuss different approaches used to build such datasets employed in NLP research.

\paragraph{Social psychology studies:} Historically, social psychology studies have provided a rich source of societal stereotypes that have been utilized to develop both resources and evaluation strategies for AI models~\cite{caliskan2017semantics}. These studies can contribute societal grounding regarding how a stereotype is perceived~\cite{fiske1991social,kite2022psychology}, as well as provide extensive examples of prevalent stereotypes about different groups~\cite{borude1966linguistic} that have been used in NLP evaluations \cite{bhatt-etal-2022-contextualizing}, and even lead to fine-tuning existing stereotype content models to LLM setting \citep{nicolas2024taxonomy, nicolas2024directionality}.

\paragraph{Crowdsourcing studies:} 
NLP researchers have recently began adapting social-psychological resources to build NLP evaluation datasets for stereotypes at scale. 
Approaches such as StereoSet~\cite{nadeem2021stereoset} and CrowsPairs~\cite{nangia2020crows} addressed the need for scaling stereotype data via crowdsourcing platforms such as Mechanical Turk. 
This crowdsourced data, while exceptionally valuable, is often tied to recognizing stereotypes reflected in specific modalities (e.g., recognizing whether a particular text reflects a stereotype), and not as a stand-alone list of social stereotypes as societal knowledge. As a result, the number of identities and unique stereotypes captured in such resources tend to be relatively small.

\paragraph{Media crawling:} Crowdsourced data, while exceptionally valuable, is often restricted in its media form (primarily text), representation (who participates in crowdsourcing), and time (reflecting a specific moment). Researchers, therefore, turned to ``big data'' resources (e.g., social networks, and web crawls) which offer a broader range of content, perspectives, and temporal data.  Existing media content, whether text, images, or videos, is shown to reflect the stereotypes present in the society. Wikipedia, for instance, documents the origins of some well-known stereotypes and describes their provenance. News articles and social media can propagate stereotypes as expressed by their authors. A popular approach for collecting such stereotypes is to crawl resources and capture co-occurrences of identity terms and attributes \citep{sap2019social,bhatt-etal-2022-contextualizing,bourgeade2023multilingual}.

\paragraph{Model-generation-based studies:} While crowdsourcing and social media based curation increase the scale of stereotype resources, they are still limited in coverage of identities and range of associated stereotypes.
More recent approaches have looked into leveraging large language models to expand coverage of stereotypes in a rapidly scalable manner and create a resource with broader coverage. When coupled with human annotations, these approaches provide validated resources that even significantly overcome selection bias of data creators
~\cite{jha-etal-2023-seegull,bhutani-etal-2024-seegull}. While this expands the state of stereotype resources across identity axes, languages, and cultures,
such an approach holds only when models are exposed (via their training data) to such social information in specific languages and about particular identity groups; thus leaving gaps in coverage across the world and many marginalized groups who are not well-represented in the online discourse.

\paragraph{Community-engaged studies:} Marginalized communities, who face some of the most severe stereotypes, are often not represented in most resources that are sourced by the previously mentioned methods. Representation is often influenced by how much these communities are written about, who gets to participate as an annotator or crowd worker, and the limits of participation in any of these roles \cite{birhane2022power}. To circumvent these gaps, recent work has engaged with underrepresented and underserved communities in a targeted manner to bridge the gaps in salient stereotype resources (e.g., \citet{alemany2022methodology,dev-2023-building,acion2023tool}). 

These approaches often offer complementary strengths and weaknesses \cite{dev2023building}. For instance, social psychological studies and community sourced studies tend to generate relatively smaller resources, but they bring forth richer and nuanced perspectives such as the perceiver group, and the extent of marginalization of the target group, while filling gaps about communities that are underrepresented in existing resources. 

\subsection{Gaps in Current Approaches}
\label{sec: gaps}
While the variety of approaches for collecting stereotypes do overlap and address some gaps (e.g., scalability and coverage), significant limitations persist across many of the mentioned approaches.

\paragraph{Stereotypes evolve over time:} Stereotypes are not static but rather temporally variable. They are influenced by how terms get reclaimed and change in meaning, historical events that lead to a shift in sentiment toward groups of people, and more (e.g., \cite{garg-etal-2018-word}). Yet, most resources capture stereotypes as a snapshot without capturing their evolving nature. For a resource to be operationalizable in bias mitigation or data and model evaluations, temporal grounding is critical. This helps resolve questions regarding factuality (e.g., French kings in 1600s being White is factual and not stereotypical) and misinformation (current Pope is not female, or Asians being associated with COVID 19 post the pandemic~\cite{lin-etal-2022-multiplex}), identification of offensive slurs or pejorative terms (e.g., the word \textit{Protestant} was derogatory in 1500s but is simply a descriptor of religious identity now) and prevalent discriminatory practices (e.g., fraction of women who could vote in the United States before and after the women’s suffrage movement~\cite{garg-etal-2018-word}).

\paragraph{Siloed Stereotype Evaluations:} 
Stereotypes affect humans and social interactions. With stereotypes reflected in generative models, they consequently impact human-AI interactions with the potential to cause a range of harmful or unpleasant effects. However, evaluations of stereotyping happen predominantly at the model checkpoints rather than at downstream use cases or applications in everyday life. They are also considered as an evaluation pillar of its own without considering the implications on various other representational or allocational harms \cite{barocas2017problem,shelby2023harms}. 

\paragraph{Lack of Consistent Conceptualization:}
As discussed by \citet{blodgett-etal-2021-stereotyping} in a thorough assessment of a number NLP measurements of stereotypes, benchmarks do not always rely on solid conceptualizations of stereotypes. Definitions of stereotype often lack critical components such as power dynamics and consistency in defining social categories. Moreover, even thorough considerations during conceptualization are not guaranteed to be accurately reflected into operationalization. While these gaps are often hard to completely eliminate, it is important to articulate them to further focus on more effective operationalizations.

\paragraph{Perceiver as a missing piece of the puzzle:} While stereotypes are born as interactions between social groups, one being the group that is perceiving and one the group that is being perceived, most frameworks and benchmarks do not consider the perceiver group and solely focus on the target group. Notably, \citet{jha-etal-2023-seegull} point out that individuals in different geographic regions are familiar with different non-overlapping stereotypes about the same identities.
While computational work on stereotypes have expanded the participant pool through crowdsourcing---although the intention for this is often to reduce the cost and time, and not to diversify the sample, they still do not take the crowdworkers background information into account in how these resources are used.

\paragraph{Lack of Contextual/Societal Grounding:} Not every over-represented association is a stereotype. Stereotypes require societal grounding for identification of harms caused \cite{bhatt-etal-2022-contextualizing,zhou-etal-2023-cobra}. Large-scale model evaluations for stereotypical or ``biased'' behavior without contextual grounding merely calibrate model tendencies.
A common example is racial bias and specifically anti-African-American stereotypes that are prevalent in the United States and rooted in colonial history, but are not similarly prevalent in South Asia where skin color does not correlate with race or nationality. Grounding a stereotype with what specific socio-cultural settings it is common in, helps build better evaluation paradigms and generative AI systems \cite{sambasivan2021re}.

\paragraph{Multilingual and Multi-Cultural settings} Stereotypes are often erroneously considered as absolute, intransient features of society that translate perfectly through languages and cultures. This however has been noted to be objectively incorrect \cite{cuddy2009stereotype}, with distinct stereotypes existing in different geo-cultures~\cite{malik-etal-2022-socially,bhatt-etal-2022-contextualizing}, some of which are expressed with words that are salient in only one language~\cite{bhutani-etal-2024-seegull}.

\begin{figure*}
    \centering
    \includegraphics[width=0.6\textwidth]{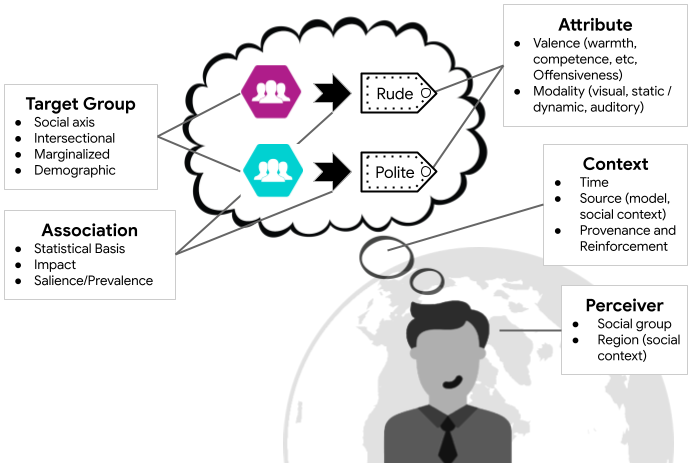}
    \caption{Framework for operationalizing stereotypes.}
    \label{fig:framework}
\end{figure*}

\section{Framework}
\label{sec: framework}

Typically, stereotypes generalize certain social groups with specific traits that allude to their agency (Competence), experience (Warmth), and often even their Morality. This is rooted in the underlying cognitive process of \textit{categorizing}, which helps humans make sense of the world by allowing them to track and distinguish others while using only a small amount of cognitive resources. We build on the social psychological conceptualization of stereotypes to introduce a framework for formalizing and depicting the content of a stereotype. Our framework is composed of five main components: the \textbf{target group}, the associated trait or \textbf{attribute}, the \textbf{association} between the target group and the attribute, the \textbf{perceiver} who holds the stereotypical belief, and the \textbf{context} in which this stereotype gets its meaning. Figure \ref{fig:framework} summarizes this framework. We now describe each of these five components below.

\paragraph{Target Group} 
The cognitive process of categorizing encourages people to think in terms of ``us (in-group) vs. them (out-group),'' which in turn leads to stereotyping.  The out-group, or \textit{target group}, is fundamental to stereotype research \cite{allport1954nature} and an integral component of a stereotype which can be characterized with the following features:
\begin{itemize}
    \item \textit{Social axis.} 
    In a social setting what separates individuals from out-groups is their perceived membership in social groups along some social axes (e.g., race, gender, ethnicity). As stereotypes are shaped by societal power structures and historical contexts, understanding the target group's socio-demographic axes helps uncover the factors that contribute to the formation and perpetuation of stereotypes. Not all social groups may be determined in terms of demographic attributes (e.g., one may hold stereotypes about \textit{techies}, or workers in the technology sector, a social group defined in terms of occupation).
    \item \textit{Intersectional.} Theories of social categorization explain that perceiving an individual as a member of multiple groups (either considered as the perceiver’s in-groups or out-groups) leads to specific stereotypes beyond the ones associated with either of the constituent groups.  The perceiver’s judgment might change when they categorize the target into in-group gender but out-group race, as opposed to out-group gender and race. So whether the group is intersectional or not is an important aspect to capture.
    \item \textit{Marginalized.} If a social group is historically marginalized, stereotypes are more likely to result in more harm. This is not to say that stereotypes toward non-marginalized groups are harmless, rather, the mechanism of harm varies based on whether or not the group is historically marginalized. This may result in discriminatory hiring practices enabled by AI systems magnifying stereotypes about temperament and suitability for employment about women and African-Americans who are known to have been marginalized in the US~\cite{bertrand-2004-labor-market,chen-2023-ethics}. Capturing such historical marginalization may help determine (and prioritize) the appropriate course of action once stereotypes are detected in model output.
    \item \textit{Demographic.} A social group can be defined by demographic features such as race, gender, or age, or other extrinsic or acquired attributes such as profession or lifestyle. Non-demographic groups may be more fluid and self-selected, whereas demographic groups are based on fixed or inherent characteristics. Stereotypes about demographic groups are often intertwined with social dynamics and can be associated with systematic discrimination. Therefore, it is important to capture this distinction \cite{crenshaw2013demarginalizing}.
\end{itemize}

\paragraph{Attribute} The \textit{attribute} describes the beliefs, assumptions, features, sentiments, or perceptions that are widely associated with members of the target group. Our conceptualization of the attribute as the characteristic associated with the target group draws heavily from the SCM \cite{fiske2018model, cuddy2007bias}.
While the association of these attributes to the target group is core to the notion of stereotypes, the attributes themselves can be characterized with certain features:

\begin{itemize}
    \item \textit{Valence.} We directly borrow valence from the SCM; the valence of the attribute can include aspects such as the associated perceived offensiveness~\cite{jha-etal-2023-seegull}, warmth, competence~\cite{nicolas2021comprehensive}, or morality~\cite{fiske2018model} of the term. The perceptions of attributes as such, and what motivates people to use them, is discussed in social psychology and NLP literature and can inform practices that rely on human ratings for identifying stereotypes. The valence of attributes may also help NLP practitioners prioritize debiasing efforts (e.g., focusing on stereotypes with offensive attributes).
    \item \textit{Modality.} Attributes manifest in different ways across different modalities. For instance, attributes like ``soft spoken'' or ``intelligent'' can be expressed clearly in text, video, or audio, but less likely to be depicted in images. On the other hand, the markers of  ``poverty'' can be vastly different in text (e.g., descriptions of poverty) versus image or video (e.g., dusty streets as visual markers of poverty that are not often verbalized). Capturing this nuance is crucial to operationalize such large databases of socio-cultural information into robust model or data interventions.
\end{itemize}

\paragraph{Association} The target group and the associated attribute together constitute the core unit of the \textit{stereotypical association}. The association itself can be characterized by the following features:

\begin{itemize}
    \item \textit{Statistical Basis (cf. Accuracy)}. The distinction between whether an association is a stereotype or factual/definitional is often blurry. For example, while it is true that Hindus often pray in temples, and this association is statistically accurate, generalizing all Hindus as temple-goers can be perceived as stereotyping, as Hinduism (like any religion) in practice encompasses a wide range of rituals beyond temple worship. On the other hand, certain associations may be readily accepted as stereotypes, but also have statistical basis: for instance, some occupational stereotypes found in NLP models align with actual US census data on job distribution \cite{garg-etal-2018-word}. 
    \item \textit{Impact}. The impact of associating an attribute to a particular group can be distinct from the attribute's valence in isolation. As such, the same attribute can have varying impacts when associated with different target groups. For example, \textit{dominating} or \textit{bossy} can be seen as slightly offensive, but when stereotypically associated with women, it pertains to professional behavior and competence and can be highly offensive. The impact captures the potential negative result of the association on the target group, distinct from (and orthogonal to) the valence of the attribute.
    \item \textit{Salience or Prevalence}. The salience or prevalence of the association can be described in various levels. It is useful to distinguish them at least at two levels from an NLP perspective: (1) \textit{model/data/language salience} represents how frequently or prominently the association appears in the model or dataset in a given language and can be measured in different ways~\cite{bhatt-etal-2022-contextualizing,jha-etal-2023-seegull}. Model salience can further be an indicator of how likely it is to influence model generations. (2) \textit{social salience} captures how widespread an association is in society, captured either at a global level, or variations across regions and communities. 
\end{itemize}

\paragraph{Perceiver} The stereotype is held by a group of people or a section of society, who we refer to as \textit{perceivers} \cite{turner1979social}. By including perceivers into this framework, we acknowledge that stereotypes are not simply properties of target groups but are actively constructed and applied by perceivers---a concept similar to the role of \textit{speaker} in NLP research \cite{hovy-yang-2021-importance}.  The socio-economic standing of this group of people, and the fraction of the population they account for are significant aspects that contribute to the severity of the stereotype.
\begin{itemize}

\item \textit{Social Group.} The social group that the perceivers belong to is crucial in understanding stereotypes because it significantly influences how they distinguish in-groups from out-groups and consequently perceive and interact with the target group. 
It is also important to note that any implications of social group membership of perceivers will differ from those of the target group's social axes. For instance, whether or not a target group is historically marginalized may be crucial in determining how stereotypes about them may be prioritized in certain contexts, but whether the the perceiver group was historically marginalized or not may not hold the same weight. 

\item \textit{Region/Social context.} Social groups often have different levels of power and status in society. This power differential can also influence how stereotypes are formed and perpetuated. Therefore the interaction of the perceivers' social group and the target group is meaningful in this context. This dynamic is an important factor for determining the possible harmfulness of the stereotype. 
\end{itemize}

\paragraph{Context}

Finally, it is crucial to remember that stereotypes are not universal or static.  They exist within specific social, cultural, and temporal contexts that shape human behaviors \cite{lewin1951intention}. Instead of implying that stereotypes speak about ``society'' in general, it is important to pinpoint both the time period and the specific reference/artifact (a dataset, a model, a geo-cultural region, etc.) that reflects the societal views in question. For instance, the perceived social norms and support for prejudice reduction in a given context can influence whether people express prejudiced attitudes \cite{devine1995racial}.  This precise component will help prevent generalizations and ensures a more accurate analysis of stereotypes.

\begin{itemize}
    \item \textit{Time}. Stereotypes are dynamic associations, reflecting shifts in social group interactions, cultural norms, and historical events over time. The perceivers' exposure to the evolving information, therefore, alters their existing stereotypes. This is an important aspect to capture in how we operationalize stereotypes in NLP research.
    \item \textit{Reference}. Stereotypes captured in NLP datasets and models, exist within specific socio-cultural contexts. Their prevalence may vary depending on which slice of society is captured in any specific dataset or model. Hence, it is important to also capture this referential context---i.e., which societal context, and which artifact, whether data or model.
    \item \textit{Provenance and Reinforcement.} The origin of a stereotype can denote the intent or purpose of reinforcing this belief on a social level. Stereotypes may be rooted in social policies, propaganda, myths or scientific misconceptions. Understanding whether a stereotype originates from scientific, religious, media, or political propaganda may be helpful for evaluating its social impact.
\end{itemize}

It is important to also note that the features in the framework may interact with one another. For example, \textit{Christians} are a minority group in India and can be seen as marginalized, whereas, the same group is not similarly marginalized in the US. 
This difference influences how stereotypes about the same target group may be dealt with in India vs. the US \cite{Kulkarni2023revisiting}.

\begin{table*}[]
    \centering
    \tiny
\begin{tabular}{p{1cm}lccccclcccccc}\toprule
\multicolumn{1}{l}{}                          & \multicolumn{5}{c}{\textbf{Target Group}}                                                                                                                                                              && \multicolumn{4}{c}{\textbf{Attribute}}                                                                                                                                            && \multicolumn{2}{c}{\textbf{Perceiver}}                                                                              \\\cline{2-6}\cline{8-11}\cline{13-14}
                                                                  &                                  &                                        &                                         &                                         &                                        &                                  && \multicolumn{3}{c}{\textbf{Valence}}                                       && \multicolumn{1}{c}{}                                        & \multicolumn{1}{c}{}                                  \\\cline{9-11}
\multirow{-2}{*}{\textbf{Source}}                                 & \multirow{-2}{*}{\textbf{Token}} & \multirow{-2}{*}{\textbf{Social Axis}} & \multirow{-2}{*}{\textbf{Int.}} & \multirow{-2}{*}{\textbf{Marg.}} & \multirow{-2}{*}{\textbf{Demo.}} && \multirow{-2}{*}{\textbf{Token}} & \textbf{Warm.} & \textbf{Compet.} & \textbf{Off.} && \multicolumn{1}{c}{\multirow{-2}{*}{\textbf{Social Group}}} & \multicolumn{1}{c}{\multirow{-2}{*}{\textbf{Region}}} \\\hline
                                                                  & Palestinian                      & nationality                            & F                                   & T                                    & T                                   && aggressive                       & low                   & high                      & high                   &         & Middle-eastern & Middle East                                                     \\
                                                                  & Netherlanders                    & nationality                            & F                                   & F                                   & T                                   && blunt                            & -                    & high                      & low                                   &                                                 &   European         &  Europe                                                      \\
\multirow{-3}{*}{SeeGULL}                                         & Afghans                          & nationality                            & F                                   & T                                    & T                                   && violent                          & low                   & high                      & high                    &                                &                                               South-Asian              &    South Asia                                                   \\\hline
{\color[HTML]{0000EE} }                                           & dentists                         & profession                             & F                                   & F                                   & F                                  && weird                            & -                    & -                        & low                                                      &                                       & -                                                            &       -                                                \\
{\color[HTML]{0000EE} }                                           & asians                           & race                                   & F                                   & F                                & T                                   && elegant                          & -                    & -                        & low                     &                                       &  -                                                           &   -                                                    \\
\multirow{-3}{*}{StereoLMs} & millennials                      & age                                    & F                                   & F                                   & T                                   && nostalgic                        & -                    & -                        & low                                           &                                       &                                      -                       &                                        -               \\\hline
                                                                  & brahmins                         & caste                                  & F                                   &  F                   & T                                   && vegetarians                      & -                    & -                        & low                         &                                       &                                        Indian                     &                      India                                 \\
                                                                  & dalits                           & caste                                  & F                                   & T                                    & T                                   && uneducated                       & -                    & low                       & high                            &                                       &            Indian                                                 &                    India                                   \\
\multirow{-3}{*}{SPICE}                                           & punjabis                         & region                                 & F                                   & F                                   & T                                   && fearless                         & -                    & high                      & low                                                          &                                       &                         Indian                                    &           India                                            \\\hline
                                                                  & old                              & age                                    & F                                   & F                                   & T                                   && fat                              & -                    & -                        & high                                           &                                       &                                        -                     &                             US                          \\
                                                                  & native Americans                 & race                             & F                                   & T                                    & T                                   && lazy                             & -                    & low                       & high                            &                                       &       -                                                      &              US                                         \\
\multirow{-3}{*}{CrowsPairs}                                                & schizophrenia                    & disability                             & F                                   & F                                   & F                                  && stupid                           & -                    & low                       & high                                      &                                       &                                -                             &                                     US                  \\\hline
                                                                  & gay men                          & SO, gender             & T                                    & T                                    & T                                   && disgusting                       & -                    & -                        & high                            &                                       &                                  -                          &                      US and Canada                                 \\
                                                                  & women                            & gender                                 & F                                   & F                                   & T                                   && objects                          & -                    & low                       & high                                            &                                       &                                   -                          &                                      US and Canada                    \\
\multirow{-3}{*}{SBF}                              & immigrants                       & nationality                            & F                                   & T                                    & F                                  && primitive                        & -                    & low                       & high                                          &                                       &                                               -              &                  US and Canada                \\\bottomrule                       
\end{tabular}
\caption{The table shows instances of stereotype from five NLP resources -- SeeGULL \cite{jha-etal-2023-seegull}, Stereotypes in LMs  \cite[StereoLMs;][]{choenni-etal-2021-stepmothers}, SPICE \cite{dev-2023-building}, CrowsPairs \cite{nangia2020crows}, and Social Bias Frames \cite[SBF;][]{sap2019social} -- imported into our framework.}
\label{tab:dataset-instances}
\end{table*}

\section{Roadmap for Operationalization}

The framework presented above is intentionally broad, with the aim of capturing all aspects of stereotypes that may be relevant in responsible AI evaluations. There may be crucial considerations that help when it comes to operationalizing the framework in specific contexts. In this section we provide such a roadmap for implementation and utilizing the framework. 

\subsection{Recommendations for Implementation}
\label{sec:recommendations}

Our framework is conceptual in nature, and is not tied to any particular implementation approach. A simpler implementation, for instance, using spreadsheets or relational databases, may suffice if the evaluation context is narrowly scoped. Table~\ref{tab:dataset-instances} shows one such tabular form implementation of our framework, where we mapped instances from five  stereotype resources that are prominently used in NLP. We chose approximately 20 examples from each of the datasets, and mapped the existing information in those datasets onto our framework. This exercise revealed cases where certain features are not applicable (e.g., \textit{vegetarianism} as an attribute does not lend itself to the SCM categories of Warmth and Competence, as it is based on a religious practice. It also revealed cases where existing datasets lack certain relevant information; e.g., StereoLMs dataset does not capture perceiver information, whereas SeeGULL and SPICE capture regional information of perceivers.

While such a simplistic implementation may suffice for demonstrative purposes, and for small scale evaluations, most real-world scenarios will require a more sophisticated implementation that can account for interrelationships between various elements of the framework. In particular, a Knowledge Graph-based implementation might be especially appropriate in this case, as it will support 
sophisticated analytics for robust data exploration and visualization, a high level of expressiveness to capture complex contextual and metadata details, adaptability to accommodate evolving insights about stereotypes, and extensibility to incorporate related entities and information from other resources.

Knowledge Graphs allow for flexible data modeling \cite{angles2017foundations}, which is crucial for capturing the evolving nature of stereotypes and their associated attributes \cite{deshpande-etal-2022-stereokg}. They emphasize relationships, enabling modeling complex relationships \cite{paulheim2017knowledge} between stereotypes and other components such as social groups. Knowledge Graphs also enable capturing nuanced knowledge about context, such as time, locale, and source provenance associated with stereotypes. Their semantic capabilities enable automated reasoning and insights, with structures suited to complex queries, visualization, and pattern detection \cite{hogan2021knowledge}. Knowledge Graphs support rapid data retrieval and efficient scaling, aided by query optimization techniques like partitioning and indexing \cite{angles2017foundations}, making them ideal for downstream mitigation efforts.

\subsection{Utilizing the Framework}
\label{sec:utilizing}

In this section, we outline some of the ways in which our framework bridges many of of the gaps identified in Section \ref{sec: gaps}. Depending on the use case, researchers should be able to identify which of the mentioned gaps might impact their conceptualization of stereotypes.  For instance, if an evaluation is aimed to be applied in a monolingual, monocultural setting, then the geo-cultural specification on stereotypes' context may not be crucial in that case. 

\paragraph{Identifying Stereotype Categories:}

Our framework goes beyond modeling stereotypes as simple relationships between an identity (e.g., Mexicans) and an attribute (e.g., lazy), and enable richer evaluations:

\begin{itemize}
    \item \textbf{Metadata utilization.} One of the  highlights of our framework is that it includes metadata that can be used to identify societal stereotypes according to specific criteria. For instance, in addition to being able to extract specific stereotypes (e.g., (\textit{Mexican}, \textit{lazy}), our framework enables us to retrieve categories of stereotypes that are of similar type (e.g., other attributes similar in meaning to \textit{lazy}). This will not only enable robust evaluation, but also identify and efficiently fill gaps in existing resources.
    
    \item \textbf{Targeted evaluation.}
    Our framework can facilitate verifying whether model responses contain specific categories of stereotypes. For instance, one might be interested in stereotypes involving identities related to a particular social axis, such as race, religion, or nationality where the identity might be that of the target group or the perceiver; stereotypes where the target is a marginalized group; stereotypes that are particularly offensive in some context; stereotypes that are prevalent in a particular culture and/or region; and more. A unified framework lends itself to such comprehensive and targeted evaluations.
\end{itemize}

\paragraph{Assessing Stereotype Evolution:}

Our framework provides a powerful lens through which we can examine the dynamic nature of stereotypes and their evolution across time and contexts.

\begin{itemize}
    
    \item \textbf{Temporal evolution analysis.} The temporal dimension in our framework allows us to track how the prevalence, valence, and/or social groups associated with stereotypes have changed over time. For instance, it was shown that gender stereotypes have evolved over time~\cite{garg-etal-2018-word}, with newer stereotypes emerging in different periods of time. Similarly, through evaluation of stereotypes and associated offensiveness, general trends of perception of different groups of people can be determined.

    \item \textbf{Contextual evolution analysis.} Stereotypes also differ across societal contexts, such as rural versus urban areas, or in different countries and cultures. This contextual evolution analysis can be uniquely conducted with a framework that not only unifies all prevalent stereotype data but also includes additional structured information regarding the perceiver, the marginalization of the target group, and more.

\end{itemize}

\paragraph{Assessing Perceivers and Context:}

Beyond simply identifying stereotypes, our framework enables a deeper exploration of how these stereotypes are shaped by and impact different perceivers and social contexts.

\begin{itemize}
    \item \textbf{Differences}. We can analyze stereotypes associated with a particular group according to different perceivers. This might be useful to understand how groups along a given spectrum may perceive a certain relevant group to gauge deeper concerns that perceivers might have about the target. For instance, we could compare the stereotypes held by Democrats and Republicans in the US toward certain groups of people, such as immigrants, trans people, or atheists.
    
    \item \textbf{Societal impact.} Stereotypes can have broader implications on society such as discrimination, inequality, or social exclusion. A unified framework enables analyzing impact in a holistic manner, tying to downstream harms \cite{shelby2023harms}.

    \item \textbf{Policy impact.} Governance policies can intervene on how technologies attenuate or exacerbate social issues such as stereotypes. Analysis of large scale impact of stereotypes in society can in turn enable impact on policies developed to protect communities and mitigate harms. Additionally, unified stereotype frameworks can enable analyzing the impact of policies on societal change~\cite{Curto2022ANO}.

    \item \textbf{Generalization.} Stereotype tuples are often studied in isolation without their linguistic context. This separation makes it impossible to fully assess the implications of different types of generalizing language \cite{davani2024genil}. Specifically, effectively identifying harmful language requires understanding the intent behind a generalization, which can range from mere mentioning a bias to actively evoking and promoting it.

\end{itemize}

\paragraph{Preventing Siloed Evaluations with Stereotype Interdependency:}

To fully grasp the complexity of stereotypes, it is crucial to move beyond isolated analyses and consider how different stereotypes interact and influence one another.

\begin{itemize}
    \item \textbf{Co-occurrence analysis.} Stereotypes can frequently co-occur, and magnify different aspects of marginalization, such as stereotypes about race and gender, or social class and ethnicity \cite{bond-et-al-race-gender}. Such patterns reveal important interdependencies that our framework enables us to identify in data and models, which in turn could lead to preventing harms to intersectional groups.

    \item \textbf{Conflict and Synergy analysis.} Multiple stereotypes can exist in a society such that they conflict or contradict each other, leading to social tensions (e.g., immigrants as both lazy and stealing jobs). 
    Stereotypes may also coexist and thus can reinforce or amplify one another, creating a more harmful impact, for instance, black women being stereotyped as loud and angry, can lead to workplace discrimination~\cite{motro-2021-race}.
    This framework enables analysis and aggregation of such interdependencies at local and global scales.

\end{itemize}

\paragraph{Detecting Stereotype Origin and Propagation}

Understanding how stereotypes emerge and spread is essential for developing effective interventions \cite{antypas2024words}, and our framework provides the tools for tracing these patterns.

\begin{itemize}
    \item \textbf{Influencer analysis.} Stereotypes originate at different points of time and are propagated differently. Recurring examination of resources and models over time helps identify key individuals, groups, or events that have contributed to the creation and/or evolution of stereotypes. For example, around the time of the COVID-19 outbreak and pandemic, anti-Asian sentiment and stereotypes were on the rise, which has been markedly observed~\cite{lin-etal-2022-multiplex}. Similar analysis can help understand the origin, propagation and severity of stereotypes. 

    \item \textbf{Media analysis.} The media often plays a critical role in shaping the perception of people worldwide,\footnote{\url{https://www.chicano.ucla.edu/files/news/NHMCLatinoDecisionsReport.pdf}} and in turn it also captures and reinforces perceptions of people already present in society.\footnote{\url{https://blog.google/intl/en-in/company-news/using-ai-to-study-demographic-representation-in-indian-tv/}} Analyzing media representations, such as movies, television shows, or news articles, contributes to the understanding of the formation and/or reinforcement of stereotypes.

\end{itemize}

\paragraph{Enhancing bias mitigation on NLP models:}

\begin{itemize}
    \item \textbf{Bias detection.} while common datasets can be used for detecting specific stereotypes in models and text, our framework enables detection on various levels, for example, using our framework, researchers could analyze a large corpus of news articles to detect the prevalence of stereotypes associating marginalized ethnic groups (target group) with offensive words (attribute) within the context of immigration debates (context). This allows for targeted analysis of bias concerning a specific marginalized group within a specific context.
    \item \textbf{Bias mitigation.} our framework enables more structured bias mitigation by only focusing on stereotypes with specific tones and levels of harmfulness and impact. Suppose our framework analysis reveals that a language model frequently generates sentences associating black women (intersectional target group) with being emotional (attribute, potentially negative valence and low Competence) in the context of workplace interactions. A bias mitigation strategy could then be designed to specifically target and reduce the frequency of these associations in the model's output, while perhaps being less concerned with other, less harmful stereotypes.
    \item \textbf{Explainability.} The framework can be used to explain the biased behavior of NLP models. For example, if a model makes a biased prediction, the framework can help to identify the underlying stereotypes that might be contributing to the bias.
    \item \textbf{Data Augmentation.} The framework can be used to generate counter-stereotypical examples for data augmentation, which can help to improve the robustness and fairness of NLP models. Furthermore, the framework can reveal missing information in datasets, for instance showing that a dataset does not include any information about perceivers or lacks data on intersectional groups. 
\end{itemize}

\section{Discussion}

Our framework provides a structured language and ontology that helps the NLP community bridge the gap between social psychological theory and computational operationalization. By forcing the explicit articulation of components like the Perceiver and Context, our model moves stereotype analysis beyond simple (\textit{Target}, \textit{Attribute}) tuples. This shift is critical for developing more granular and robust evaluation methodologies that are sensitive to socio-historical nuances. For instance, classifying a bias as merely ``racial'' is insufficient; a proper evaluation requires specifying the relationship—who is holding the belief (Perceiver) about whom (Target) and in what geo-cultural setting (Context)—to determine the appropriate mitigation strategy. Furthermore, a structure like this is essential for building interdependent stereotype knowledge bases that support complex analytical queries, paving the way for the next generation of context-aware and culture-sensitive debiasing techniques in LLMs.

We provided recommendations for implementing the framework using Knowledge Graphs in Section \ref{sec:recommendations}, however, we also acknowledge that depending on the specific use case in which stereotypes need to be operationalized, developers might not find it efficient to incorporate all aspects of the framework in their design; for instance, the operational complexity and lack of scalability and the role of human oversight in maintaining such a Knowledge Graph introduce significant costs to a project. 
Thus Section \ref{sec:utilizing} discusses how different research and technical problems benefit from specific aspects of the framework.  We also acknowledge the need for research into more computationally lightweight alternatives for implementation that still preserve the framework's richness, allowing smaller research teams or production systems to adopt its core principles without incurring high maintenance costs.

The current framework provides the \textit{what} (the components of a stereotype), but future work must integrate the \textit{how}---specifically, developing methods to parse and encode the linguistic context (e.g., sarcasm, metaphor, active vs. passive voice) that modulates a stereotype's expression and potential for harm. Future efforts should alos rigorously test and adapt the framework's components to demonstrate utility in a broader range of NLP tasks beyond LLM evaluation, such as bias detection in knowledge distillation or fairness in multimodal generation. This would solidify the framework's value as a universal tool for responsible AI.

\section{Limitations}
While our framework captures various aspects of stereotyping by drawing from social psychology and NLP, we acknowledge its potential limitations. First, our goal is for the framework to improve stereotype evaluation and mitigation in LLMs. This inherent focus on model-centric applications and the subjectivity in interpreting the application can limit the generalizability of the framework to other NLP tasks. Second, while our framework emphasizes the essential role of Context in shaping stereotypes, we recognize that context is inherently multifaceted and dynamic, encompassing a vast array of factors, including but not limited to social norms, historic events, individual experiences, and power dynamics. Due to this complexity, any attempt to model the context is inevitably incomplete. Instead, we encourage researchers to explicitly consider and document the relevant contextual factors in their efforts, even if those factors expand beyond the specific elements included in the current framework. Moreover, several studies in NLP tend to the linguistic context in which stereotypes are expressed and explore nuanced communication elements such as linguistic modalities, reasons, motivations, sarcasm, and parody as they co-occur with stereotyping language. A focused linguistic effort is essential for incorporating such linguistic factors with the core aspects of stereotypes discussed in this paper.
Therefore, ongoing critical engagement and reflection is highly necessary for linguistic, social and historical grounding of stereotype evaluations.

\bibliography{anthology,custom}

\appendix
\section{Glossary}

\begin{description}
    \item[\textbf{Stereotype}]  --- A cognitive generalization about a specific social group, often consisting of widely shared beliefs and assumed traits associated with its members.
    \item[\textbf{Categorizing}] --- The fundamental cognitive process of grouping objects, events, or people into categories, which is essential to the formation of stereotypes.
     \item[\textbf{Intersectionality}] --- The concept that individuals belong to multiple social groups simultaneously, and that stereotypes targeting these intersectional identities create unique forms of bias beyond those of the component groups.
    \item[\textbf{Stereotype Content Model (SCM)}] --- A foundational social psychological framework that posits group stereotypes vary along two primary, universal dimensions: Warmth and Competence.
    \item[\textbf{Agency-Beliefs-Communion Model (ABC) }] --- A theoretical extension of the SCM that adds Beliefs as a third dimension, alongside refined aspects of Agency (Competence) and Communion (Warmth).
    \item[\textbf{Warmth}] --- The SCM dimension that captures perceived good or ill intent, reflecting traits like friendliness, sincerity, and morality.
    \item[\textbf{Competence}] --- The SCM dimension that captures perceived capability or status, reflecting traits like intelligence, skill, and agency.
    \item[\textbf{Perceiver}] --- The individual, group, or section of society that holds and applies a specific stereotypical belief about the target group.

\end{description}

\end{document}